\documentclass[preprint,preprintnumbers,amsmath,amssymb,superscriptaddress]{revtex4}
\usepackage{txfonts}% Physical Review B
\usepackage{graphicx}% Include figure files
\usepackage{dcolumn}% Align table columns on decimal point
\usepackage{bm}% bold math
\usepackage{array}

\begin{document}

\preprint{APS/123-QED}

\title{Orbital Order and Spin Nematicity in the Tetragonal Phase of Electron-doped Iron-Pnictides NaFe$_{1-x}$Co$_{x}$As}

\author{R. Zhou}
\author{L. Y. Xing}
\author{X. C. Wang}
\author{C. Q. Jin}
\affiliation{Beijing National Laboratory for Condensed Matter Physics,\\
Institute of Physics, Chinese Academy of Sciences, Beijing 100190, China}
\author{Guo-qing Zheng}
\affiliation{Beijing National Laboratory for Condensed Matter Physics,\\
Institute of Physics, Chinese Academy of Sciences, Beijing 100190, China}
\affiliation{Department of Physics, Okayama University, Okayama 700-8530, Japan}
\date{\today}% It is always \today, today,
             %  but any date may be explicitly specified

\begin{abstract}
{In copper-oxide and iron-based high temperature (high-$T_{\rm c}$) superconductors, many physical properties exhibit in-plane anisotropy, which is believed to be caused by a rotational symmetry-breaking nematic order, % \cite{Lawler,Daou,Chu,Shen,Kasahara,Zhou,Dai-neutron},
whose origin  and its relationship to  superconductivity remain elusive. %is under  scrutiny. % \cite{Chu2,Fernandes_PRB,Lv,Lee,Onari,Kontani_2014,Fernades_np}.
In many iron-pnictides, a tetragonal-to-orthorhombic structural transition temperature $T_{\rm s}$ coincides with  the magnetic transition  temperature $T_{\rm N}$, %\cite{Kasahara,Zhou,Dai-neutron},
making the orbital and spin degrees of freedom highly entangled.
NaFeAs is a system where  $T_{\rm s}$ = 54  K is well separated from  $T_{\rm N}$ = 42 K, which helps simplify the experimental situation. %disentangle different degrees of freedom. % \cite{NaFeAs}.
Here we report nuclear magnetic resonance (NMR)  measurements on NaFe$_{1-x}$Co$_x$As (0 $\leq x \leq$ 0.042) that revealed orbital and spin nematicity  occurring at a temperature $T^{\rm *}$ %(as high as 90 K)
far above  $T_{\rm s}$  in the tetragonal phase.
We show that %the observed
the NMR spectra splitting and its evolution  can be explained by %is due to
an incommensurate orbital order that sets in below    $T^{\rm *}$ and becomes commensurate below $T_{\rm s}$, which  brings about the observed spin nematicity.} %, which explains why the electronic nematicity  exists above $T_{\rm s}$.}
\end{abstract}

\pacs{74.70.Xa, 74.25.nj, 74.25.-q, 75.25.Dk}

\maketitle
Understanding the normal state out of which high-$T_{\rm c}$ superconductivity (SC) develops is  an important task in condensed-matter physics. In   copper-oxide high-temperature superconductors, the normal state  %the superconducting transition temperature
%above $T_{\rm c}$
deviates from the conventional state described by Landau Fermi liquid theory. In particular, below a certain temperature $T^*$, a so-called pseudogap state emerges, breaking the rotation symmetry of the underling lattices  \cite{Lawler,Daou}.
In  iron-pnictide or iron-selenide  high-$T_{\rm c}$ superconductors, many physical properties in the normal state also show strong anisotropy (nematicity), breaking the four-fold rotation (C4) symmetry \cite{Chu,Shen,Kasahara}. %in the  competes or coexists with superconductivity \cite{Chu,Shen,Zhou}. % \cite{Lawler,Daou,Chu,Shen,Zhou}.
For example, in the parent Fe-pnictide BaFe$_2$As$_2$, electronically-driven nematicity was discovered in the in-plane resistivity below a tetragonal-to-orthorhombic structural transition temperature $T_{\rm s}$   \cite{Chu,Chu2}. Soon after the transport measurements, angle resolved photo-emission spectroscopy (ARPES) found that the degeneracy of the Fe-3$d_{xz}$ and 3$d_{yz}$ orbitals is lifted \cite{Shen}. %at low temperatures . %, hinting at a possible  relationship between orbital order and the anisotropy in the transport properties.
Later on, nematicity was also found in other properties ranging from magneto-elastic property in chemically-pressurized BaFe$_2$As$_2$ \cite{Kasahara}, to spin dynamics   in carrier-doped  BaFe$_2$As$_2$ \cite{Dai-neutron}, and to local electronic structure around defects %in NaFeAs
 even  above $T_{\rm s}$ \cite{Rosenthal,Marc}. % the structural transition.
Theoretically, both spin \cite{Fernandes_PRB} and orbital origin \cite{Lv,Lee,Onari,Kontani_2014} have been proposed for the cause of the experimentally-observed nematicity.
In the BaFe$_2$As$_2$ family, however, antiferromagnetism (AF)  sets in simultaneously at $T_{\rm s}$ or slightly below \cite{Zhou}. As a result, it is unclear whether the   transition is  driven by spin degree of freedom \cite{Fernandes_PRB} or by orbital degree of freedom \cite{Lv,Lee,Onari,Kontani_2014}. Neither is it clear whether the  nematicity is caused by a static  \cite{Kasahara} or a fluctuating order \cite{Dai-neutron}. % have been proposed.
Therefore,
identifying the origin of the nematicity has become an urgent  issue, since  it is believed that the interaction  leading to such a nematicity may also be responsible for
 the high-$T_{\rm c}$   superconductivity\cite{Fernades_np,Yang}.
 %In any case, since no static orbital order was detected by ARPES above $T_{\rm s}$ \cite{Feng}, it has been suggested that the driving force is
%many theory believe that the mechanism for the superconductivity is related to magnetic fluctuation\cite{Mazin,Kuroki,Scalapino}, but the appearance of nematic order shows people another possible route,
%orbital fluctuations  has also been proposed as a mechanism to achieve high temperature superconductivity \cite{Kontani}.
%Therefore, the origin of the nematic order has become an central issue in iron-pnictides \cite{Fernades_np}. %Using the strain produced by piezoelectric,

%However, due to the coupling of spin and orbit, it is hard to distinguish between such two mechanisms, and the origin of electronic nematicity is still a mystery.

%NaFe$_{1-x}$Co$_{x}$As is an electron-doped system  where
NaFeAs is a unique system where  $T_{\rm s}$ = 54  K is well above  $T_{\rm N}$ = 42 K. Only 2.7 percent of Co substituting for Fe gives rise to the maximum $T_{\rm c}$ = 21 K \cite{NaFeAs}, which makes the system a clean one with much less doping-induced disorder    than other systems.
In this Communication, we report evidence pointing toward  orbital order  at a temperature $T^*$ (as high as 90 K) that is far above $T_{\rm s}$ in NaFe$_{1-x}$Co$_{x}$As by $^{75}$As and $^{23}$Na NMR spectroscopy. %Furthermore, , the spin-lattice relaxation
We further revealed a spin nematicity in this system by the spin-lattice relaxation rate ($1/T_1$) measurements, and show that it can be understood as a direct consequence of the orbital order.

The single crystals % samples
of NaFe$_{1-x}$Co$_{x}$As used for  the  measurements were grown by the self-flux method\cite{Rosenthal}.
%A detailed description can be found elsewhere .
In order to prevent sample degradation, % by air,
The samples were covered by Stycast 1266$^{\circledR}$ in a glove box filled with high-purity Ar gas\cite{supplemental_materials}. The typical sample size is 3mm$\times$3mm$\times$0.1mm. The Co content $x$ was determined by energy-dispersive x-ray spectroscopy. The $T_{\rm c}$ was determined by DC susceptibility measured by a %superconducting quantum interference
SQUID device. The NMR spectra were obtained by integrating the spin echo as a function of  frequency at %a constant external magnetic field
$H_0$ = 11.998 T. The $T_1$ was measured by using the  saturation-recovery method, and determined by a good fitting to the theoretical curve \cite{method_T1}.  %  to $1-\frac{M\left( t \right)}{M\left( \infty  \right)}=0.9{{\exp }^{-6t/{{T}_{1}}\;}}+0.1{{\exp }^{-t/{{T}_{1}}\;}}$, where $M(t)$ is the nuclear magnetization at time $t$ after the saturation pulse .

 The $^{75}$As or $^{23}$Na nucleus with spin $I$ = 3/2 has a nuclear quadrupole moment $Q$ that couples to the electric field gradient (EFG) ${{V}_{\alpha \alpha}}$ ($\alpha =x, y, z$), relating to the %observables called the nuclear quadrupole resonance (
NQR frequency tensors ${{\nu }_{\alpha}=\frac{eQ}{4I(2I-1)}V_{\alpha\alpha}}$. % ($\alpha$ = $a$, $b$, $c$)
  Therefore, both $^{75}$As- and $^{23}$Na-NMR are good probes for a structural phase transition as shown in %the parent compound
NaFeAs  where the principal axes are along the crystal axes \cite{Kitagawa_NaFeAs}. In addition, the As site is very close to the Fe plane so that the As-$p$  and Fe-$d$ orbitals strongly hybridize, which makes $^{75}$As NMR also a  sensitive and unique probe for detecting an orbital order since a disparate occupation in As-$p$ orbitals will produce an asymmetric   EFG.

\begin{figure}
\includegraphics[width = 10 cm]{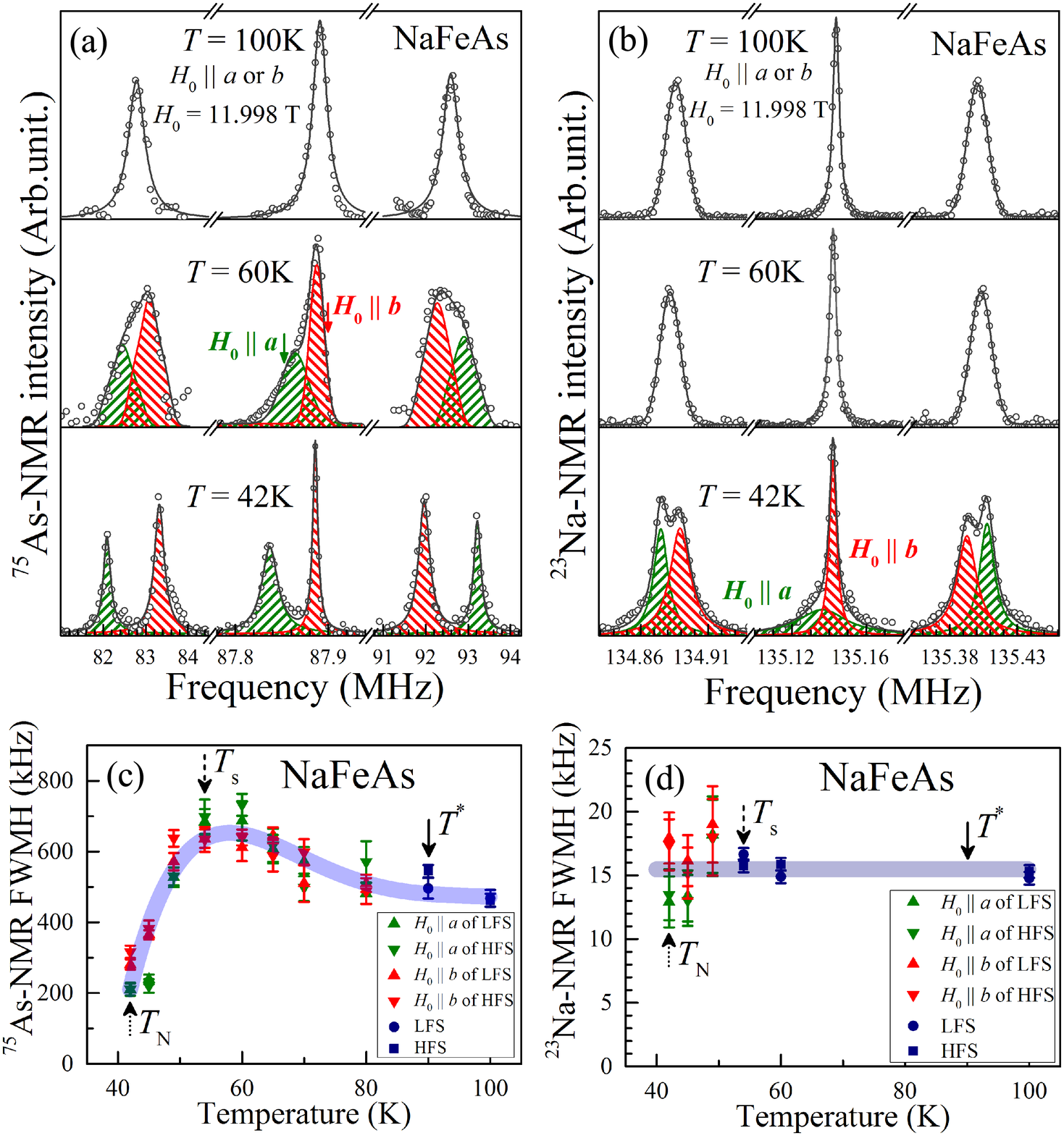}
\caption{\textbf{(a-b)}  $^{75}$As-  and $^{23}$Na-NMR spectra. % obtained by sweeping the frequency at a fixed external field $H_0$ = 11.998 T applied along the
%with $H_0 \parallel a$-axis ($b$-axis), % (Fe-Fe bond direction),
%at $T>T^*$, $T^*>T>T_{\rm s}$ and $T<T_{\rm s}$, respectively.
The three peaks at $T$ = 100 K respectively correspond to  the low frequency satellite (LFS), % $\frac{1}{2}\leftarrow\rightarrow\frac{3}{2}$ transition),
central %($\frac{1}{2}\leftarrow\rightarrow-\frac{1}{2}$)
transition and high frequency satellite (HFS).  %, $-\frac{3}{2}\leftarrow\rightarrow-\frac{1}{2}$ transition) lines. %, each of which is split at lower temperatures.
%The $^{75}$As-NMR spectra are broaden below $T^{*}$, while $^{23}$Na-NMR spectra are unchanged. This indicates that broadening in $^{75}$As-NMR spectra above $T_{\rm s}$ is only from the orbital order. The Below $T_{\rm s}$, for both $^{75}$As-NMR and $^{23}$Na-NMR spectra, the clear splitting is seen. Most remarkably, the linewidth of $^{75}$As-NMR peaks decrease with temperature decreasing, which suggests the orbital order changes from incommensurate to commensurate at $T_{\rm s}$.
The middle panel in \textbf{(a)} shows the simulation of 2D-incommensurate orbital order model for $T$ = 60 K spectra. Green (red) shadow area represents the transitions with  $H_0 \parallel$ $a$-axis ($b$-axis).
%The grey line indicates the difference of the value from second order quadrupole shift between $a$ and $b$-axis.
%The spectra are fitted by only one Lorentz function above $T^{*}$, but two Gaussian functions between $T^{*}$ and $T_{\rm s}$, and two Lorentz functions below $T_{\rm s}$.
The green and red arrows show the positions at which 1/$T_{1a}$ and 1/$T_{1b}$ was measured, respectively. At $T$ = 100 K, 1/$T_{1}$ measured at  low- and high-frequency tails gives rise to the same value.
\textbf{(c-d)}, $T$-dependence of the full width at half maximum (FWHM) of each peak.}
\label{spectra}
\end{figure}

When a magnetic field $H_0$ is applied along  $i$-axis ($i$ = $a$ or $b$), %(Fe-Fe bond direction),
the NMR resonance frequency $f$ is expressed by \cite{Abraham}
\begin{equation}
\label{f}
\begin{aligned}
&{{f}_{m\leftrightarrow m-1,i}}={{\gamma }_{N}}{{H}_{0}}\left( 1+{{K}_{i}} \right)+\frac{1}{2}{{\nu }_{c}}\left( m-\frac{1}{2} \right)\left( {{n}_{i}}\cdot \eta -1 \right)+{{a}_{m}}\delta {{f}_{i}}
\end{aligned}
\end{equation}

, where % $\gamma_{N}$ is the nuclear gyromagnetic ratio,
$K_i$ is the Knight shift, %means the magnetic field is applied along $i$-axis.
 $m$ = 3/2, 1/2 and -1/2, and $n_i$ = $\mp$ 1.
%The ${{\nu }_{\alpha}=\frac{eQ}{4I(2I-1)}V_{\alpha\alpha}}$ ($\alpha$ = $a$, $b$, $c$) is the nuclear quadrupole resonance frequency tensor, with $c$-axis being its principal axis in NaFe$_{1-x}$Co$_x$As.
%(see below for detailed identification of $n_i$).,
$\eta \equiv \frac{\left| {{V}_{xx}}-{{V}_{yy}} \right|}{{{V}_{zz}}}\text{=}\left|\frac{ {{\nu }_{a}}-{{\nu }_{b}} }{{ {\nu }_{a}}+{{\nu }_{b}}} \right|$ is the  asymmetry parameter of the EFG, which measures a nematicity in the $ab$-plane.
Finally,
%, and by spin-lattice relaxation rate (see below) we believe that $n_a$ = - 1 and $n_b$ = 1.
 ${{a}_{m}}\delta {{f}_{i}}$ is the second-order quadrupolar shift when $H_0$ is applied parallel to $i$-axis, and is given as,
\begin{equation}
\label{f2}
 \begin{aligned}
  & \delta {{f}_{a}}=\frac{{{\left( {{v}_{b}}-{{v}_{c}} \right)}^{2}}}{12\left( 1+{{K}_{a}} \right){{\gamma }_{N}}{{H}_{0}}}=\frac{3v_{c}^{2}}{16\left( 1+{{K}_{a}} \right){{\gamma }_{N}}{{H}_{0}}}{{\left( 1-\frac{\eta }{3} \right)}^{2}} \\
 & \delta {{f}_{b}}=\frac{{{\left( {{v}_{a}}-{{v}_{c}} \right)}^{2}}}{12\left( 1+{{K}_{b}} \right){{\gamma }_{N}}{{H}_{0}}}=\frac{3v_{c}^{2}}{16\left( 1+{{K}_{b}} \right){{\gamma }_{N}}{{H}_{0}}}{{\left( 1+\frac{\eta }{3} \right)}^{2}}
\end{aligned}
\end{equation}
This %second-order quadrupolar
correction only needs to be considered for the central transition ($m$=1/2) line, so $a_{1/2}$ = 1 and $a_{3/2}$ = $a_{-1/2}$ =0.
%
%For the central transition line, the frequency $f$ is related to $K_i$ and $\delta {{f}_{i}}$ which depends on $\eta$, while for the satellite peaks ($m$ = 3/2 and -1/2), $f$ only depends on $\eta$.
%Meanwhile, the EFG comes from both the on-site charge density distribution determined by the electronic orbits and the surrounding ions which is sensitive to  the crystal structure.
For a material with C4 rotation symmetry, $\eta = 0$. However,  $\eta > 0$
 if  C4  symmetry is broken. %, the EFG tensor along $a$ or $b$-axis is not identical, so that .
Therefore,  for a twined single crystal with C2 symmetry,  the  field configurations of  ${{H}_{0}}\parallel $ $a$-axis and ${{H}_{0}}\parallel $ $b$-axis will give a different  ${{f}_{m\leftrightarrow m-1, i}}$, leading to a splitting of both satellite and central peaks.

Figure \ref{spectra} shows the evolution of  the % $^{75}$As-NMR and $^{23}$Na-
NMR spectra in NaFeAs. At high temperature ($T$), only one central transition %line
and a pair of satellites are observed.
Below $T^{*}$ = 90 K,  a broadening of both central and satellite lines
was seen in the $^{75}$As-NMR spectra, but not in  the  $^{23}$Na-NMR lines. % (middle panel of Fig. \ref{spectra} (a) and (b)).
% Each peak can be well reproduced by assuming two split peaks, which indicates that they are formed by two broad peaks.
With further decreasing $T$, all the  $^{75}$As-NMR lines become narrower  below $T_{\rm s}$, and a clear splitting is observed. % (lower panel of Fig. \ref{spectra} (a)).
Same is true for $^{23}$Na-NMR lines  below $T_{\rm s}$. % (lower panel of Fig. \ref{spectra} (b)).
Note that $T^{*}$ is much higher than the $T_{\rm s}\sim$ 54 K confirmed by both previous neutron scattering measurement \cite{NaFeAs_Dai} and our resistivity data \cite{supplemental_materials}.

\begin{figure}
\includegraphics[width= 7 cm]{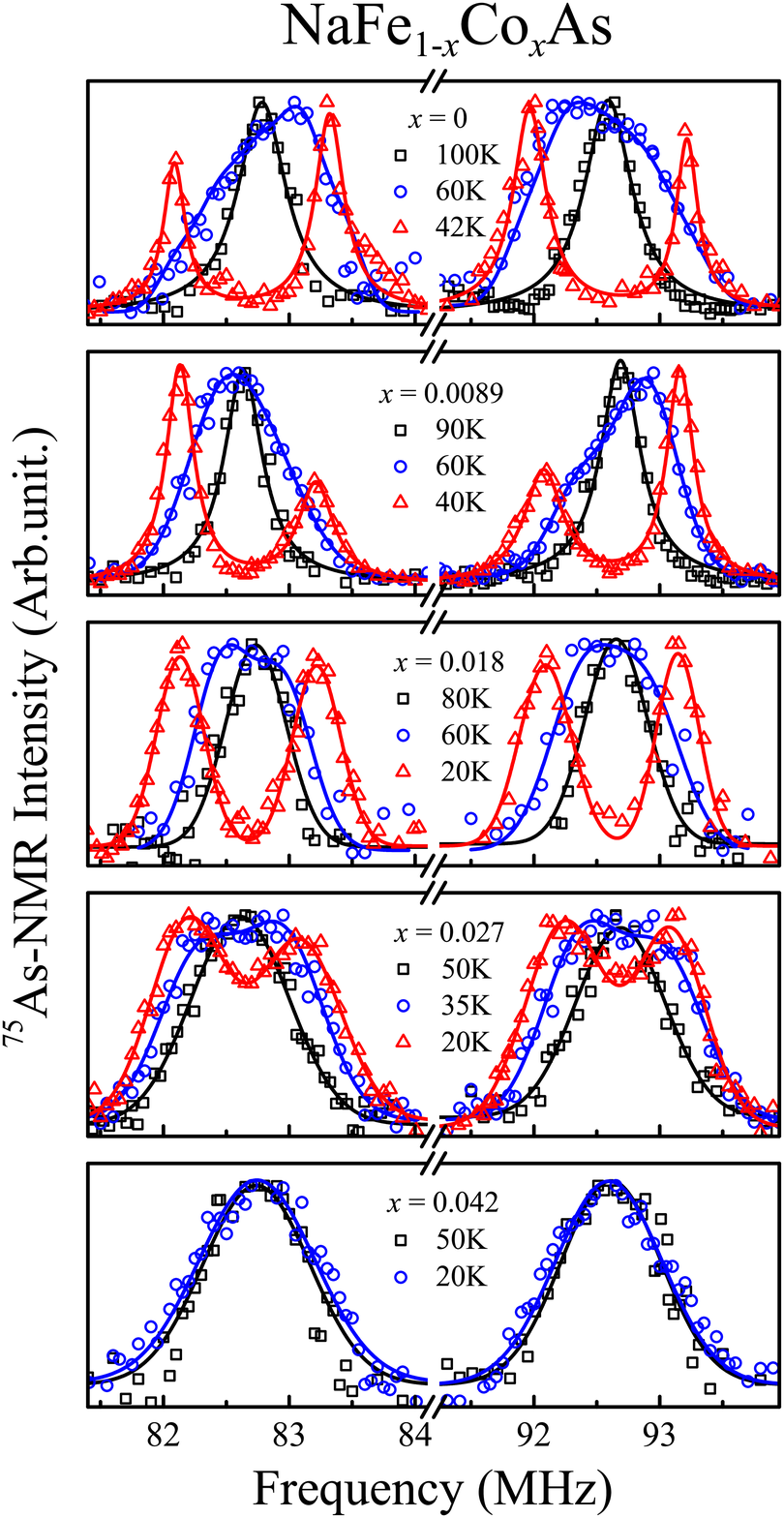}
\caption{  $T$-evolution of the $^{75}$As-NMR satellite peaks. %\textbf{a-d}, The satellite peaks with the external field $H_0$ = 11.998 T applied along the $a$-axis or $b$-axis above and below $T^{*}$ and $T_{\rm s}$ for $x$ = 0, 0.0089, 0.018, 0.027, respectively.
For $x$ = 0, 0.0089,  0.018 and 0.027, the  peaks are broadened below $T^{*}$, and split below $T_{\rm s}$. %For $x$ = , the clear change can still be observed below 45K. However, the linewidth of the splitted peaks is much larger, because the orbital order is still incommensurate at $T_{\rm c}$.
For $x$ = 0.042, however, no clear change in the spectrum is detected down to $T$ = 20 K.
%The spectra are fitted by only one Lorentz or Gaussian function above $T^{*}$, but two Lorentz or Gaussian functions below $T_{\rm s}$.
}
\label{spectra_all}
\end{figure}
%There are two possibilities that can account the  $^{75}$As-NMR lines splitting far above $T_{\rm s}$.

One apparent possibility for the  $^{75}$As-NMR lines broadening (splitting) %far above $T_{\rm s}$
is that there are some small local orthorhombic domains existing above $T_{\rm s}$ formed by tiny uniaxial pressure from disorders \cite{Marc,Inoue} or uniaxial strains due to  epoxy encapsulation.
However, this can be  ruled out since $^{23}$Na-NMR spectra do not change below $T^{*}$ and there is no angular  dependence of $T^{*}$ \cite{supplemental_materials}.
The other is that  orbitals order in the real tetragonal phase. % without the help of structural distortion.
In this case, the origin of EFG asymmetry % $\eta_{As}$  %disparity of the occupation of As-4$p_x$ and As-4$p_y$ orbitals.
%The
is the Fe orbital order parameter $\Delta$. For example, for the orbital splitting  found in ARPES \cite{Shen}, one can write $\Delta \propto$ ($n^{3d}_{xz}-n^{3d}_{yz}$), where $n$ is the electron density. It will  produce a population disparity between As-4$p_x$ and 4$p_y$, ($n^{4p}_{x}-n^{4p}_{y}$), through Fe-As orbital hybridization.  Such  disparity  was explained by electronic mechanism \cite{Fernandes_PRB,Lv,Lee,Onari,Kontani_2014}, as well as by local-density approximations  calculation \cite{Shimojima_LDA}. At the moment, we  cannot rule out other form of $\Delta$ that can produce a finite ($n^{4p}_{x}-n^{4p}_{y}$).
The As-NQR frequency tensor $\nu_{x,y,z}$ is related to $n^{4p} _{x,y,z}$ as \cite{ZhengGQ}
\begin{equation}
\label{nu}
  \left[\begin{array}{c}
    \nu_{x} \\
    \nu_{y} \\
    \nu_{z} \\
  \end{array}\right]
  =\nu_{0}
  \left[\begin{array}{c}
    n^{4p}_{x}-\frac{n^{4p}_{y}+n^{4p}_{z}}{2} \\
    n^{4p}_{y}-\frac{n^{4p}_{x}+n^{4p}_{z}}{2} \\
    n^{4p}_{z}-\frac{n^{4p}_{x}+n^{4p}_{y}}{2} \\
  \end{array}\right]
\end{equation}
where $\nu_0$ is the NQR frequency when there is one electron (hole) in each 4$p$-orbital. It follows that
%\begin{equation}
%\label{nu-eta}
% \begin{aligned}
$\left| \nu_{x}-\nu_{y} \right| = \frac{3\nu_0}{2}\left| n^{4p}_{x}-n^{4p}_{y}\right|$
%\end{aligned}
%\end{equation}
 , therefore  $\eta_{As}\propto \left| n^{4p}_{x}-n^{4p}_{y} \right| \propto \Delta$.% $\eta_{As}\propto \left| n^{4p}_{x}-n^{4p}_{y} \right| \propto \Delta$. % measures the difference in the population of the As-4$p_{x}$ and  As-4$p_{y}$ orbits, $\Delta n$, due to Fe-3$d$ orbital ordering.
% \begin{equation}
%\label{eta}
% \begin{aligned}

%\end{aligned}
%\end{equation}

%In our study, we found that  The Na site is much far away from Fe plane\cite{supplemental_materials}, and the Na electron orbits do not directly overlap with the orbits of Fe. Therefore Na is not sensitive to the orbital order, but still sensitive to the structure transition. Thus, our results rule out the existence of orthorhombic domains.
%Therefore the change of $^{75}$As spectra in Fig. \ref{spectra} above $T_{\rm s}$ can only be due to an orbital order.
%In BaFe$_2$(As$_{1-x}$P$_x$)$_2$ system, magnetic torque measurements has revealed an electronic nematicity at $T^{*}$ that is much higher than the normal structural transition $T_{\rm s}$, but the structure there also breaks the $C4$ symmetry instantly\cite{Kasahara}.
%Therefore, it has been widely believed that the structure can not remain tetragonal when an orbital order forms.

\begin{figure}
\includegraphics[width=11 cm]{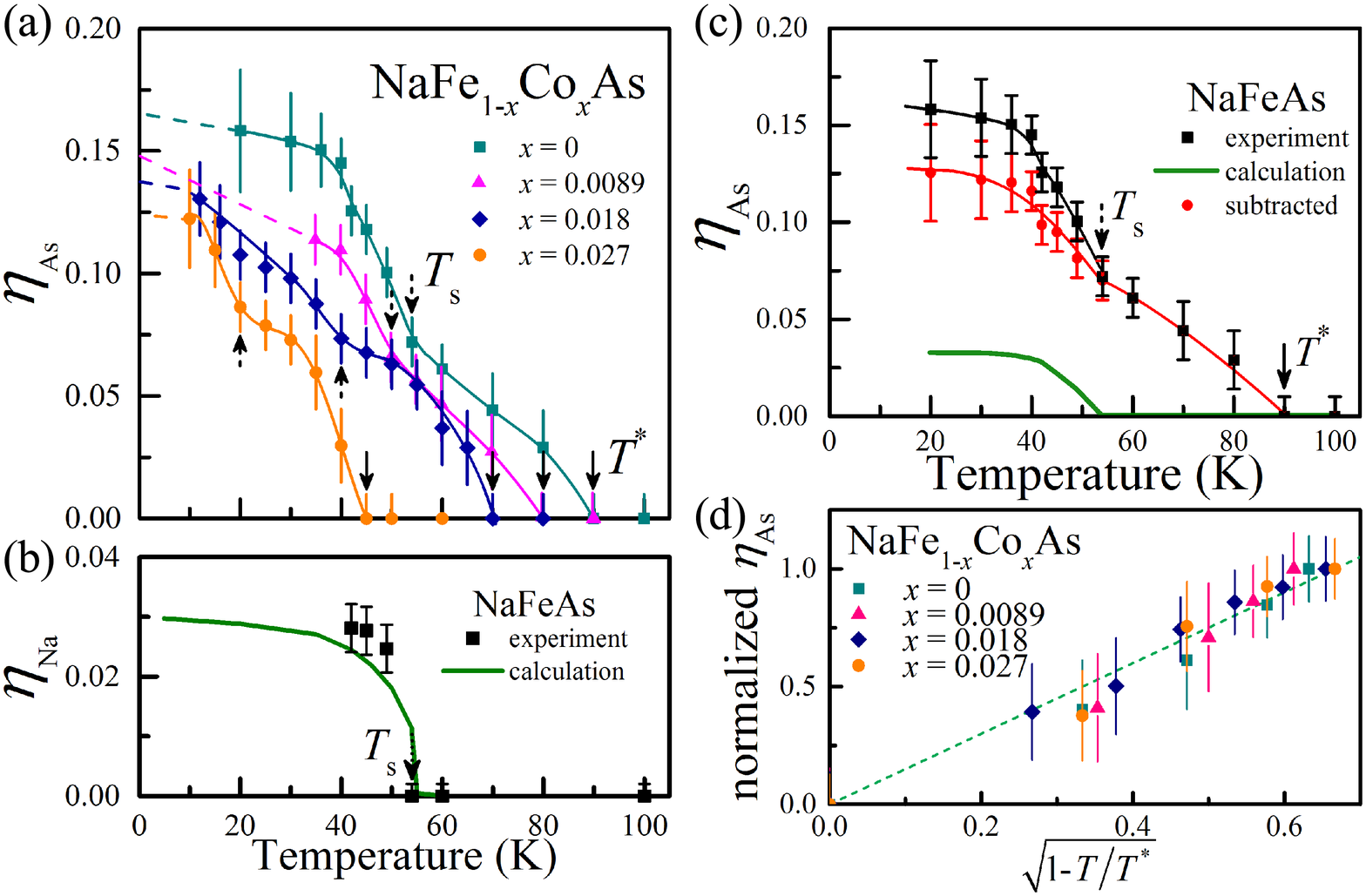}
\centering
\caption{The evolution of the EFG asymmetry parameter $\eta$. \textbf{(a)} $T$-dependence of  $\eta_{\rm As}$  % at the As site
for various $x$. The solid and dotted arrows indicate $T^*$ and $T_{\rm s}$, respectively.
%The open olive diamond is the calculation value of ${\eta}_{As}$, by using the lattice distortion value from former XRD study\cite{Parker_XRD}.
%For x = 0, no change is seen around $T_{\rm N}$ = 42 K, but a kink is observed at $T_{\rm s}$ = 54 K which is due to the structural transition. %anisotropic spin fluctuation.
%For x = 0.0089 and 0.018, since the spin fluctuation will become weaker with doping, the kink will become more and more indistinct.
%Most remarkably, for x = 0.027, instead of suppression, we observed the enhancement of $\eta$ below $T_{\rm c}$, which indicates that the superconductivity and orbital nematicity do not compete with each other.
\textbf{(b)} $T$-dependence of $\eta_{\rm Na}$ %at the $^{23}$Na-site
for $x$=0.  The green curve is the contribution due to the structure change obtained %by the first principle calculation which  was performed
by all-electron full-potential linear augmented plane wave  method \cite{Blaha,Oguchi}, by using the lattice parameter from neutron scattering  \cite{NaFeAs_Dai} and X-ray diffraction \cite{Parker_XRD}.
%spectra and the lattice distortion from XRD data\cite{Parker_XRD}, which obviously shows that between $T_{\rm s}$ and $T^{*}$, the structure remains tetragonal. The solid olive square represents the calculation ${\eta}_{Na}$, which is in good agreement with experimental data.
\textbf{(c)} Experimental, calculated and the subtracted data of $\eta_{\rm As}$   for $x$=0.
\textbf{(d)} $\eta_{\rm As}$ normalized by its value at $T_{\rm s}$ against $\sqrt{1-{T}/{{{T}^{*}}}\;}$ for various $x$. %A linear relationship is seen  in the vicinity of $T^{*}$. %0.65, above which  %proportional to $\sqrt{1-{T}/{{{T}^{*}}}\;}$ below $T^{*}$. %This suggests $\eta$ is Landau-type order parameter of second-order transition.
% a deviation from the linearity is observed,  due to the increasing of $\eta$ below $T_{\rm s}$.
}
\label{yita}
\end{figure}

Below we show  that an
%By analyzing the NMR spectra, we find that the orbital order is
incommensurate orbital order in the tetragonal phase, which becomes commensurate below $T_{\rm s}$,  can consistently account for the observed results.
%for $T^* \geq T \geq T_s$ but becomes commensurate for $T\leq T_s$, as elaborated below.
%
%The situation is different in the present case where the orbital order above $T_{\rm s}$ is not the same as the one below $T_{\rm s}$. As seen in Fig. \ref{spectra} (c) and (d), the full width of half maximum (FWHM) of both low frequency satellite (LFS) and high frequency satellite (HFS) of $^{75}$As-NMR spectra increase below $T^{*}$. Yet no change is seen in $^{75}$Na-NMR spectra. Considering that the broadening around $T_{\rm s}$ is about 200 kHz, which is much larger than the linewidth of central peak(less than 50 kHz at 54K), this means that the broadening is not due to a distribution of the Knight shift. Moveover, below $T_{\rm s}$, the FWHM rapidly decreases with temperature decreasing. Furthermore, the broadening of the spectra due to impurity, if any should not decrease with decreasing temperature. All these indicate that the broadening of $^{75}$As-NMR spectra is owing to the appearance of a distribution of $\eta$.
%
The observed behavior is very similar to the crossover of commensurate to incommensurate antiferromagnetic order in NaFeAs\cite{Kitagawa_NaFeAs}.
Generally speaking, in a commensurate density-wave state, the NMR line  reflects the small number of physically non-equivalent nuclear sites in the unit cell so that the linewidth is small. In an incommensurate state, however, since the translational periodicity is lost, %and as a result,
 the number of non-equivalent nuclear sites is larger which gives rise to a larger linewidth \cite{Blinc}. %Such a difference was well known in charge-density wave states .
%Figure 3 illustrates the orbital ordered states for the present case.
%In the one-dimensional, commensurate charge-density wave state, a characteristic line shape with two sharp edges are observed \cite{Blinc}.
%In the present case, however,  no sharp edge was found, which  %which has been seen in incommensurate antiferromagnetic order\cite{Kitagawa_NaFeAs}.
% can be explained by a two-dimensional incommensurate ordered state.
In the present case,
A modulation due to orbital order  will cause an additional term in  %spatial distribution in
the  resonance frequency at As site ($x,y$). Let this term  be a cosine function as $\left[ \cos \left( \frac{2\pi }{a}{{q}_{x}}\cdot x + {\theta_{x}} \right)+\cos \left( \frac{2\pi }{b}{{q}_{y}}\cdot y + {\theta_{y}} \right) \right]$,
%
%\begin{equation}
%\label{2D-IC}
% \begin{aligned}
%{{f}^{\text{2D}}}_{m\leftrightarrow m-1,i}\left( x,y \right)={{f}_{m\leftrightarrow m-1,i}}+\Delta {{f}_{m\leftrightarrow m-1,i}}\left[ \cos \left( \frac{2\pi }{a}{{Q}_{x}}\cdot x + {\theta_{x}} \right)+\cos \left( \frac{2\pi }{a}{{Q}_{y}}\cdot y + {\theta_{y}} \right) \right]
%\end{aligned}
%\end{equation}
%
%\begin{equation}
%\label{2D-yita}
% \begin{aligned}
%  \eta \text{=}{{\eta }_{0}}\text{+}\Delta {{\eta }_{\text{x}}}\cos \left( \frac{2\pi }{a}{{Q}_{x}}\cdot x+{{\phi }_{a}} \right)+\Delta {{\eta }_{y}}\cos \left( \frac{2\pi }{b}{{Q}_{y}}\cdot y+{{\phi }_{b}} \right)
%\end{aligned}
%\end{equation}
%
where
%$a$ and $b$ are lattice parameters, and they are the same above the structural transition.
$q_x$ %, ${\phi }_{a}$
and $q_y$ %, ${\phi }_{b}$
are the two-dimensional (2D) wave vectors %. and phase along $a$ and $b$-axis, respectively. Here, we assume it is the complete incommensurate state and the phase has not $Q$ dependent. ${{f}_{m\leftrightarrow m-1,i}}$ is the same as Eq. \ref{f}.
%  $\Delta {{f}_{m\leftrightarrow m-1,i}}$ is the modulation amplitude
and $\theta_{x,y}$ is the phase. Then,
for commensurate order, %because it is ferro-orbital order\cite{Shen}, the period of the orbital order is the same as lattice, meaning that
%$q_{x} = q_{y}$ = 1 ( $\frac{q_{x} \cdot x}{a}$ or $\frac{q_y \cdot y}{b}$ becomes an integer). So that
the additional term  becomes %${{f}^{\text{2D}}}_{m\leftrightarrow m-1,i}\left( x,y \right)={{f}_{m\leftrightarrow m-1,i}}+
%$\Delta {{f}_{m\leftrightarrow m-1,i}}
$\left[ \cos  {{\theta }_{x}} \text{+}\cos  {{\theta }_{y}}  \right]$, which is site-independent.
For incommensurate order, however, this term  is site-dependent, which leads to a broadening of the spectrum.
% we do not need to consider the exact value of wave vector and phase.
By convoluting % the frequency distribution
%\begin{equation}
%\label{2D-f}
% \begin{aligned}
%${{I}_{m\leftrightarrow m-1,i}}\left( f \right)=\sum\limits_{x,y}{\int{\delta \left( f-f_{m\leftrightarrow m-1.i}^{2\text{D}}\left( x,y \right) \right)df}}$
%\end{aligned}
%\end{equation}
with a Gaussian  function %and took the parameters from the spectra from the fitting of 45 K spectra whose orbital order is already commensurate
 \cite{supplemental_materials}, we can reproduce the spectra as shown in %the middle panel of
Fig. \ref{spectra} (a).
In passing,we note that $T^*$=90 K for  $x$=0  is consistent with the temperature below which scanning tunneling microscope  found local electronic nematicity by quasiparticle interference \cite{Rosenthal}. %
%However, it should be emphasized that the orbital order we inferred %found here
%is not induced by external strain \cite{Rosenthal} or pinned by defects \cite{Rosenthal,Marc}, since, in that case, the local crystal structure around the defects break C4 symmetry and should result in a broadening of Na-NMR spectra above $T_{\rm s}$.
%, which is not seen as shown in Fig. \ref{spectra}.
%In our case, by Na-NMR, we excluded the possibility of local distortion, indicating what we observed is a more intrinsic orbital order when structure still has C4 symmetry.

 %The calculated spectra are only broadened with no sharp edge.
%
%So we suggest that the orbital order is incommensurate above $T_{\rm s}$, and it is more close to two dimensional type.
%

Below $T_{\rm s}$, the $^{75}$As-NMR spectra become narrower and each peak is well resolved. Moreover, no NMR intensity loss is observed below  $T_{\rm s}$ ord $T^{*}$\cite{supplemental_materials}. All these imply that all the As sites have the same environment. That is, the orbital order becomes commensurate.
%So we can observe the decrease of linewidth with temperature decreasing as shown in Fig. \ref{spectra} (c). Above $T_{\rm s}$, the spectra did not show too many details, we cannot obtain the real modulation function.
%But if the incommensurate order is more close to two dimensional type, the change of each split line from commensurate to incommensurate will only be the broadening. Therefore we can fit it with two Gaussian functions and obtain $\eta_{As}$.
%
The doping dependence of the spectra is  shown in Fig. \ref{spectra_all}. %The same
As for $x$=0, a peak splitting was also found above $T_{\rm s}$ for $x$ = 0.0089,  0.018 and 0.027, which get well resolved at $T_{\rm s}$. % for $x$ = 0.0089 and  0.018.  %For the optimal doping ( $x$ = 0.027 ), a spectral change was also observed below $T^{*}$ = 45 K. By $^{23}$Na-NMR we observed the broadening below 20K. However, the spectrum will naturally broaden below $T_{\rm c}$ (under 12 T is around 20 K), so we could not conclude whether the structural distortion also occurs in optimal doping sample\cite{supplemental_materials}.
%
%
%that this sample has no structural distortion
 For $x$ = 0.042, however, no change of the spectra was found down to $T$ = 20 K.

%We confirmed that it is the incommensurate orbital order, and spin nematicity is also found by spin-lattice relaxation 1/$T_1$ below.

%In order to investigate the origin of the electronic nemacticity, we  studied the physical properties of the nematic order.
More quantitative data are shown in Fig. \ref{yita} where
%The change in the character of the orbital order at $T_{\rm s}$  is  supported by
the evolution of  $\eta$ is demonstrated.
 %  shows the temperature dependence of $\eta_{As}$ and $\eta_{Na}$, respectively.
%For all Co-doping contents,
The $\eta_{As}$ develops continuously below $T^{*}$, showing a saturation tendency approaching $T_{\rm s}$.
In contrast,  $\eta_{Na}$ shows up only below $T_{\rm s}$ and the absolute value is much smaller than  $\eta_{As}$, indicating that it is purely due to the structural transition. %being consistent with the second-order nature of the phase transition.
There are two contributions to the observed $\eta$, $\eta$= $\eta_{lattice}$ + $\eta_{orbital}$, where $\eta_{lattice}$ is  due to surrounding lattice and $\eta_{orbital}$ is due to orbital order on Fe site.
By the first principle calculation, we find that the observed $\eta_{Na}$ is well explained by the change in $\eta_{lattice}$; the discrepancy is about 10\%.
 Another remarkable feature of  $\eta_{As}$ is that  it  increases steeply again below $T_{\rm s}$, which cannot be accounted by %.
 %For $T_{\rm s}<T<T^{*}$, %the asymmetry parameter
% One obvious additional contribution to $\eta_{As}$ below $T_{\rm s}$ is that due to symmetry breaking of the structure.
the  %first-principle
calculated $\eta_{lattice}$. % (green curve in Fig. \ref{yita} (d)). % shows that the lattice change cannot account for  the observed increase.
%which changes the EFG from the nearby As sites and also influences the orbital order. This will increase $\eta$.
%As shown in Fig. \ref{yita} (b),   the calculated $\eta$ due to structure transition at Na sites  is in good agreement with experimental data.
%However, the calculated increase of $\eta$ at As sites cannot   %is less than 1/3 of the observed one,
%which suggests  that the steep increase of $\eta_{As}$ below $T_{\rm s}$ is due to a change in the character of the orbital order.
The red circles in   Fig. \ref{yita} (d) is  the net increase after subtracting the effect due to the lattice change.  A clear kink can be seen at $T_{\rm s}$, which is true even after multiplying the calculated result by a factor of 1.1$\sim$1.15.
The increase is consistent with %e steep increase of $\eta_{As}$ below $T_{\rm s}$ can be understood by
the incommensurate-to-commensurate transition.
%
%
%we subtract the $\eta_{cal}$ from experimental data for NaFeAs, as shown in supplementary materials. We can still clearly observe a kink at $T_{\rm s}$, indicating that structural symmetry change is not the only reason for such behavior.
%Another possibility which could enhance $\eta_{As}$ is the incommensurate to commensurate transition.
In the incommensurate state, $\left| n^{4p}_{x}-n^{4p}_{y} \right|$ is inhomogeneous and $\eta_{As}$ probes the averaged $\left| n^{4p}_{x}-n^{4p}_{y} \right|$. %Thus the average frequency of As site equals the first part of Eq. \ref{2D-IC}.
In the commensurate state, $\eta_{As}$  measures the homogeneous $\left| n^{4p}_{x}-n^{4p}_{y} \right|$, which can be larger \cite{supplemental_materials}. % (see Supplemental Materials).

\begin{figure}
\includegraphics[width= 8.5 cm]{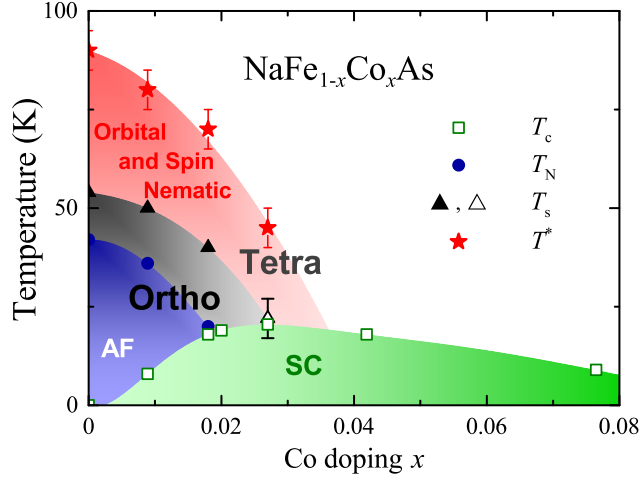}
\caption{The obtained phase diagram. % of NaFe$_{1-x}$Co$_x$As.
%The $T_{\rm N}$ and $T^{*}$ are obtained from NMR spectra.
For $x < $0.027, %such determined
$T_{\rm s}$ agrees well with that from resistivity \cite{supplemental_materials}. For $x$ = 0.027, the $T_{\rm s}$ coincides with  $T_{\rm c}$ so that  direct comparison with resistivity  is unavailable. To distinguish with other compositions, the data point is represented by an open triangle.
%AF and SC denote the antiferromagnetic  and the superconducting states, respectively.
Ortho and Tetra represent the orthorhombic and tetragonal phase, respectively.}
% The error bar represents the temperature interval in measuring the NMR spectra. } %The curves are guides to the eyes.}
\label{phase}
\end{figure}

Finally, it is worthwhile pointing out that  $\eta_{\rm As}$ shows a linear relationship with $\sqrt{1-{T}/{{{T}^{*}}}\;}$  for all samples in the vicinity of $T^{*}$, as shown in Fig. \ref{yita} (d), suggesting that the nematic order undergoes a Landau-type-like second-order phase transition. %The deviation from the linearity  above $\sqrt{1-{T}/{{{T}^{*}}}\;} = 0.65$ is due to the increasing of $\eta$ below $T_{\rm s}$.
%and the second part of Eq. \ref{2D-IC} should be considered. If the phase of the orbital order is smaller than $\pi$/2, an increase of $\eta$ will be observed.
%In order to understand the origin of the kink, we calculated structural distortion related $\eta_{cal}$ for both As and Na sites
%
The $T^{*}$ and $T_{\rm s}$ results obtained by NMR are summarized in the phase diagram  shown in Fig. \ref{phase}.

\begin{figure}
\includegraphics[width=10cm]{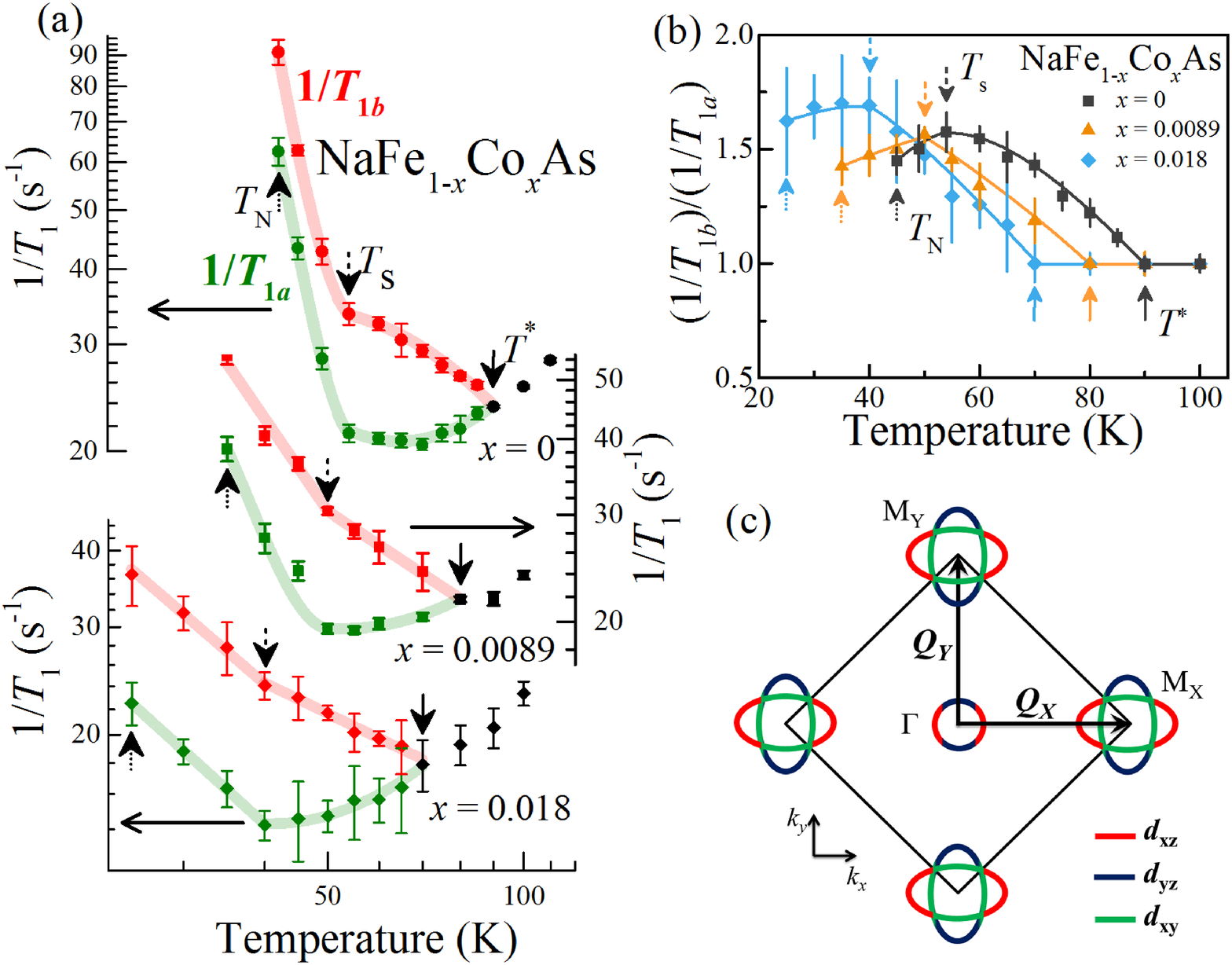}
\centering
\caption{ \textbf{(a)}  $T$-dependence of 1/$T_1$  for various $x$. %The error bar of 1/$T_1$ is the standard deviation in fitting the nuclear magnetization recovery curve.
\textbf{(b)} $T$ dependence of 1/$T_1$ ratio. % for $x$ =0, 0.0089 and 0.018. The ratio increases from unity below $T^{*}$, indicating that spin nematicity develops. % and 1/$T_1$ will become different. %With decreasing temperature, 1/$T_1$ will rapidly increase at $T_{\rm s}$, due to the enhancement of ${{\chi }''}_{a}$.
%Intriguingly, $\frac{{{\left( 1/{{T}_{1}}\right)_{b}^{\text{AF}}  }}}{{{\left( 1/{{T}_{1}}\right)_{a}^{\text{AF}}  }}}$  starts to decrease below $T_{\rm s}$, which is an unexplained feature at the present stage.
%It may be due to a development of spin fluctuations anisotropy (${{{{\chi }''}}_{a}}>{{{{\chi }''}}_{c}}$), but more work such as precise neutron scattering measurement needs to be done in the future.
%The curves are guides for the  eyes.
Solid, dashed and dotted arrows indicate $T^{*}$, $T_{\rm s}$ and $T_{\rm c}$, respectively.
\textbf{(c)} Schematics of the Fermi surface (FS) obtained by ARPES \cite{Zhang} and the spin fluctuations wave vectors. The outer FS centered at $\Gamma$ is omitted here for clarity. The color represents the Fe-3$d_{xz,yz,xy}$ orbital character.}
\label{T1}
\end{figure}

Next we turn to the spin dynamics of this system. %,  %. ,  spin nematicity
which was also clearly seen below $T^{*}$ in $1/T_1$. %the spin-lattice relaxation rate  %  in the Knight shift $K$ and .
%Figure \ref{T1} (b) displays the temperature dependence of $K$ for various samples. For all samples, the difference of $K_a$ and $K_b$ is not obvious below $T^{*}$, because of the overlap of two splitting central lines. However, below $T_{\rm s}$ it shows clear difference. As seen in Fig. \ref{spectra}, the grey line indicates the difference from the second order quadrupole shift which is smaller than the splitting of the central peaks. This indicates that the $K_a$ and $K_b$ is indeed different in the nematic order. Such behavior is due to the nematic susceptibility that is also observed in FeSe\cite{Bohmer,Baek}. For parent compound, both $K_a$ and $K_b$ decrease more quickly below $T_{\rm s}$,
% similar to the pseudogap behavior. Actually, in BaFe$_2$(As$_{1-x}$P$_x$)$_2$, ARPES data suggests that there may be some pseudogap below nematic transition\cite{Shimojima}. For optimal doping, both $K_a$ and $K_b$ decrease with temperature decreasing below $T_{\rm c}$, and we also do not observe the oblivious suppression of $K$ nematic below $T_{c}$.
Figure \ref{T1} (a) shows the  1/$T_1$ results for $x$ = 0, 0.0089 and 0.018. Below $T^*$, 1/$T_1$ measured at the positions corresponding to $H_0\parallel [100]_{\rm o}$ ($a$-axis) and $H_0\parallel [010]_{\rm o}$ ($b$-axis) shows opposite $T$-dependence. Here we assign the direction with larger NQR frequency tensor to the $a$-axis. Figure \ref{T1} (b) shows the ratio of the two $T_1$.
%
%For optimal doped sample, the change of 1/$T_1$ becomes smaller and more impurity is introduced, which strongly influence $T_1$ measurement. So we could not get the same data for $x$ = 0.027 sample.
%
%The right panel shows the 1/$T_1$ due to the contribution of antiferromagnetic fluctuations.
%Below $T^{*}$, 1/$T_{1a}$ and 1/$T_{1b}$ were respectively measured at the position marked by the green and red arrows in Fig. \ref{spectra}.
%This indicates that the spin nematicity appear instantly below $T^{*}$, as elaborated below.
As in Ba(Fe$_{1-x}$[Ni,Co]$_x$)$_2$As$_2$ \cite{Zhou,Co-NMR-prl_Ning},
%NaFe$_{1-x}$Co$_x$As is an electron-doped system,
 $1/{T}_{1}$ arises  %has two parts of contributions. % \cite{supplemental_materials}.One is
from the antiferromagnetic spin fluctuations  %${{\left( 1/{{T}_{1}}\  \right)}^{\text{AF}}}$,
%and  the other  is
and the contribution due to the intra-band  (DOS at the Fermi level),  %, ${{\left( 1/{{T}_{1}}\  \right)}^{\text{intra}}}$.
%In the present case, the contribution of spin fluctuations %${{\left( 1/{{T}_{1}}\  \right)}^{\text{AF}}}$
but the former is dominant \cite{supplemental_materials}.
%much less than ${{\left( 1/{{T}_{1}}\  \right)}^{\text{AF}}}$ in the present case.
%We did not observe an appreciable difference of $K$ below $T^{*}$ (see Supplementary Material), which means that ${{\left( 1/{{T}_{1}}\  \right)}^{\text{intra}}}$ barely changes below the nematic transition. This means that the difference between 1/$T_{1a}$ and 1/$T_{1b}$ comes mainly from the antiferromagnetic spin fluctuation.
%After subtracting the intra-band contribution (see Supplemental Materials),
%we find that  ${{{\left( {1}/{{{T}_{1}}}\; \right)_{b}}^{\text{AF}}}}/{{{\left( {1}/{{{T}_{1}}}\; \right)_{a}}^{\text{AF}}}}$ is a unity above $T^{\rm *}$, but increases below $T^{\rm *}$ , reaching to 2 at $T_{\rm s}$, %before showing a drop below,
%as shown in Fig. \ref{T1}(b)-(d).
We show below that the anisotropy of $1/T_1$ is a natural consequence of   the orbital order.

The magnetic order on the Fe atoms below $T_{\rm N}$ is of stripe type with ordering vectors $Q_X$ = ($\pi$,0)   \cite{NaFeAs_Dai}.  Above $T_{\rm N}$, however,  magnetic fluctuations from $Q_Y$ = (0,$\pi$)  also exist and have the equal amplitude as those from $Q_X$. Since As sits   above or below  the center of the  square formed by four irons,   1/$T_1$ of $^{75}$As along the orthorhombic $a$-direction %($[100]_{\rm o}$)
or $b$-direction %($[010]_{\rm o})$
sees  antiferromagnetic spin fluctuations from both $Q_X$ and $Q_Y$  as follows \cite{Li}
%, ${{\left( \frac{1}{{{T}_{1}}} \right)_{a}}^{{Q}_{X}}} \propto {{A}^{2}}{{{{\chi }''}}_{a}}$, ${{\left( \frac{1}{{{T}_{1}}} \right)_{b}}^{{Q}_{X}}} \propto {{A}^{2}}\left( {{{{\chi }''}}_{a}}+{{{{\chi }''}}_{c}} \right)$ and ${{\left( \frac{1}{{{T}_{1}}} \right)_{a}}^{{Q}_{Y}}} \propto {{A}^{2}}\left( {{{{\chi }''}}_{b}}+{{{{\chi }''}}_{c}} \right)$, ${{\left( \frac{1}{{{T}_{1}}} \right)_{b}}^{{Q}_{Y}}} \propto {{A}^{2}}{{{{\chi }''}}_{b}}$.
\begin{equation}
\label{Qx}
 \begin{aligned}
  & {{\left( \frac{1}{{{T}_{1}}} \right)_{a}}^{{Q}_{X}}} \propto {{A}^{2}}{{{{\chi }''}}_{a}} \\
 & {{\left( \frac{1}{{{T}_{1}}} \right)_{b}}^{{Q}_{X}}} \propto {{A}^{2}}\left( {{{{\chi }''}}_{a}}+{{{{\chi }''}}_{c}} \right) \\
\end{aligned}
\end{equation}
and %Likewise,  1/$T_1$ from antiferromagnetic spin fluctuation at $Q_Y$ = (0,$\pi$) is
\begin{equation}
\label{Qy}
 \begin{aligned}
  & {{\left( \frac{1}{{{T}_{1}}} \right)_{a}}^{{Q}_{Y}}} \propto {{A}^{2}}\left( {{{{\chi }''}}_{b}}+{{{{\chi }''}}_{c}} \right) \\
  & {{\left( \frac{1}{{{T}_{1}}} \right)_{b}}^{{Q}_{Y}}} \propto {{A}^{2}}{{{{\chi }''}}_{b}} \\
\end{aligned}
\end{equation}
Here $A$ is the hyperfine coupling constant and ${{\chi }''}_{j}$ ($j$ = $a$, $b$, $c$) is  the imaginary part of the staggered susceptibility. %Here we assume the hyperfine coupling constant is isotropic
%No appreciable difference in the Knight shift $K = A\cdot\chi(0)$ was observed at $T^{*}$ \cite{supplemental_materials}, which means that the hyperfine coupling $A$ does not change appreciably.
The measured ${ {\left( {1}/{{{T}_{1}}}\; \right)_{i}}}$ ($i=a,b$ ) can then be written as
%${{\left( \frac{1}{{{T}_{1}}} \right)_{i}}}={{N}_{X}}{{\left( \frac{1}{{{T}_{1}}} \right)_{i}}^{{Q}_{X}}}+{{N}_{Y}}{{\left( \frac{1}{{{T}_{1}}} \right)_{i}}^{{Q}_{Y}}}$.%should pick up contributions from both $Q_X$ and $Q_Y$ as
\begin{equation}
\label{T1H}
 \begin{aligned}
{{\left( \frac{1}{{{T}_{1}}} \right)_{i}}}={{N}_{X}}{{\left( \frac{1}{{{T}_{1}}} \right)_{i}}^{{Q}_{X}}}+{{N}_{Y}}{{\left( \frac{1}{{{T}_{1}}} \right)_{i}}^{{Q}_{Y}}}
\end{aligned}
\end{equation}
where $N_X$ ($N_Y$) is the relative weight of contribution from $Q_X$ ($Q_Y$), with $N_X$ + $N_Y$ = 1.
It then follows
%$\frac{{{\left( 1/{{T}_{1}}\right)_{b}  }}}{{{\left( 1/{{T}_{1}}\right)_{a}  }}} = \frac{{{\chi }''}_{c}+{{\chi }''}_{a}+\frac{N_Y}{N_X}{{\chi }''}_{b}}{{{\chi }''}_{a}+\frac{N_Y}{N_X}({\chi }''_{b}+{\chi }''_{c} )}$, %Therefore, the quantity ${{\left( 1/({{T}_{1}})_{a}\  \right)}^{\text{AF}}}$/${{\left( 1/({{T}_{1}})_{b}\  \right)}^{\text{AF}}}$%\begin{equation}
\begin{equation}
\label{T1R}
 \begin{aligned}
\frac{{{\left( 1/{{T}_{1}}\right)_{b}  }}}{{{\left( 1/{{T}_{1}}\right)_{a}  }}} = \frac{{{\chi }''}_{c}+{{\chi }''}_{a}+\frac{N_Y}{N_X}{{\chi }''}_{b}}{{{\chi }''}_{a}+\frac{N_Y}{N_X}({\chi }''_{b}+{\chi }''_{c} )}
\end{aligned}
\end{equation}
which
%The  quantity $\frac{{{\left( 1/{{T}_{1}}\right)_{b}^{\text{AF}}  }}}{{{\left( 1/{{T}_{1}}\right)_{a}^{\text{AF}}  }}}$ therefore
measures a change in the ratio $\frac{N_Y}{N_X}$. The $1/T_1$  ratio will always be a unity as long as  $N_X$ = $N_Y$. On the other hand, in the limit of $N_X \sim$1 and  $N_Y \sim$0, the ratio will be 2, since polarized inelastic neutron scattering  found  that the anisotropy in the low-energy spin excitations above $T_{\rm s}$ is small, if any  \cite{Song}.
%Also, polarized inelastic neutron scattering study found  that the low-energy spin excitations are isotropic, ${{{{\chi }''}}_{a}}={{{{\chi }''}}_{b}}$, above $T_{\rm s}$ in NaFeAs \cite{Song}.

As seen  in Fig. \ref{T1} (b), % $N_X$=$N_Y$ above $T^{\rm *}$ as expected.
the observed ratio $\frac{{{\left( 1/{{T}_{1}}\right)_{b}  }}}{{{\left( 1/{{T}_{1}}\right)_{a}  }}}$ increases  below $T^*$,  indicating that $N_X$ increases  and $N_Y$ decreases.
%The fact of $\frac{{{\left( 1/{{T}_{1}}\right)_{b} }}}{{{\left( 1/{{T}_{1}}\right)_{a}  }}}$ = 2 at  $T=T_{\rm s}$ implies that $N_X \sim$1 and  $N_Y \sim$0 there,
These results are a natural consequence of orbital order with the occupation of Fe-3$d_{xz}$ becoming larger than Fe-3$d_{yz}$ which changes the FS nesting condition so that spin fluctuations with $Q_X$ becomes dominant \cite{Su}.
This is because $Q_X$ connects the  Fermi pocket centered at $\Gamma$ = (0,0) with that centered at $M_X$ = ($\pi$,0) consisting of $d_{xz}$ orbital, and $Q_Y$ connects the $\Gamma$ Fermi pocket with that centered at $M_Y$ = (0, $\pi$) consisting of $d_{yz}$ orbital (see Fig. \ref{T1} (c)). Finally, we note that an anomaly is found at $T_{\rm s}$ in $1/T_1$ of both directions, which is consistent with a change in the character of orbital order, but the detailed analysis of the anomaly and theoretical explanation are a topic of future investigation. % but calls for theoretical interpretation.

Previously,  electronic nematicity was  found in  BaFe$_2$As$_2$ \cite{Shen,Kasahara,Zhou} and FeSe \cite{Shimojima_1,Baek,Watson,Bohmer,Qisi} systems, but it  occurs right at $T_{\rm s}$.  %simultaneously at the structural transition temperature.
Above $T_{\rm s}$, only  fluctuations were observed  \cite{Fernandes_PRL,Meingast_PRL}.
In Ni-doped BaFe$_2$As$_2$, although anisotropy was found in the spin susceptibility above $T_{\rm s}$, the system was under a uni-axial pressure and it was attributed to a fluctuating order \cite{Dai-neutron}. By contrast, no external driving force was applied in the present case, thus the observation of a static order at the time scale of 10$^{-8}$ sec is unprecedented.
%The evidence pointing toward a static order in the tetragonal phase%(the NMR spectrum change)

%In summary, we have presented the result of the systematic NMR measurements on single crystal samples of NaFe$_{1-x}$Co$_{x}$As. A broadening of $^{75}$As-spectra was observed far above $T_{\rm s}$. However, $^{23}$Na-NMR spectra did not change, which indicates an intrinsic orbital order is formed in the tetragonal phase. We further found that the splitting peaks becomes narrower and the EFG parameter $\eta$ becomes larger just below $T_{\rm s}$, which suggests that the orbital order in the tetragonal phase is incommensurate and becomes commensurate below $T_{\rm s}$. By measuring the 1/$T_1$, we found that the spin nematicity is also formed in the orbital order state. Our finding highlights the uniqueness of the iron-pnictide (selenides) as a multi-orbital system and shows that the field where orbital degree of freedom plays an important role is a rich research frontier that deserves more exploration.

In summary, we have presented  the systematic NMR measurements on single crystals  of NaFe$_{1-x}$Co$_{x}$As.  The $^{75}$As-spectra were
broadened at  $T^*$ far above $T_{\rm s}$ %due to a splitting of the transition peaks
and  get well split below $T_{\rm s}$.
The EFG asymmetry parameter $\eta$ emerges below $T^*$ and increases abruptly  below $T_{\rm s}$.
However, the $^{23}$Na-NMR spectra showed no change until $T_{\rm s}$.
These results can be explained by %indicate that
an  incommensurate orbital order  formed in the tetragonal
phase %, and  suggest that the orbital order in the tetragonal phase is incommensurate but
which becomes commensurate below $T_{\rm s}$.
A spin nematicity is also found below $T^*$, %by the measurement of 1/$T_1$,
which can be understood as
%was shown to be
a direct consequence of the orbital order.
\vspace{0.5cm}

\begin{acknowledgments}
We thank T. Xiang   for helpful discussion and comments, M.-H. Julien   and S. Onari for a critical reading of the manuscript, S. Maeda and T. Oguchi for advice and help in the EFG calculation, Z. Li and J. Yang for assistance in some of the measurements. This work was partially supported by CAS Strategic Priority Research Program, No. XDB07020200 and by a 973 project National Basic Research Program of China,
No. 2012CB821402.
\end{acknowledgments}

%{\bf Author contributions}
%The single crystals were grown by L.Y.X., X.C.W. and C.Q.J.. The NMR measurements were performed by R.Z. with
%supervision of G.Q.Z.. G.Q.Z performed the EFG calculation, coordinated the whole work and wrote the manuscript, which was supplemented by R.Z.. All authors have discussed the results and the interpretation.

%{\bf Additional information}
%Supplementary information is available in the online version of the paper.
%Correspondence and requests for materials should be addressed to G.Q.Z.

%{\bf Competing financial interests}
%The authors declare no competing financial interests.

%\begin{figure}
%\includegraphics[width=12cm]{orbit}
%\caption{\textbf{The sketch of the evolution of orbital order.} The top and bottom panels show the sketch of the evolution of orbital order and corresponding low frequency satellite peak. Color of the $d_{xz}$ orbit represents the difference in the occupation of the $d_{xy}$ and $d_{xz}$ orbits $\Delta n$ = $n_{d_{xz}}$ - $n_{d_{yz}}$. Solid and dashed circles depicts the As above and below Fe plane, respectively. The length of double-headed arrows indicate the value of $\eta$. In the incommensurate state, $\eta$ will have a large distribution, resulting in the broadening of the spectra. }
%\label{orbit}
%\end{figure}

\end{document}